\documentclass[aps,amsmath,amssymb,twocolumn,showpacs]{revtex4-1}

\usepackage{graphicx}
\usepackage{dcolumn}
\usepackage{bm}
\usepackage{color}

\begin{document}

\title{Optical forces and torques exerted on coupled silica nanospheres: unexpected effects due to the multiple scattering}


\author{R.~M. Abraham Ekeroth$^{1}$}

\email{mabraham@exa.unicen.edu.ar}

\affiliation{$^{1}$Instituto de F\'isica Arroyo Seco (IFAS-CIFICEN), Universidad Nacional del Centro de la Provincia de Buenos Aires, Tandil, CP 7000 Buenos Aires, Argentina}

\date{\today}

\begin{abstract} 
Optically coupled nanoparticles suffer the action of multiple electromagnetic forces when they are illuminated by light. In general, two kinds of forces are commonly assumed: binding forces that make them attract/repel each other and scattering forces that push the system forwards. Tangential forces and orbital torques can also be induced to align the dimer with the electric field. In this work, new degrees of freedom are found for a dimer of silica nanospheres under illumination with linearly-polarized plane waves. The results have a general validity for arbitrary mesoscale systems: multiple scattering of light induces unexpected torques and unbalanced forces. These torques include spin contributions to the movement of the whole system.

The results are supported by previous works and pave the way for the engineering of nanoscale devices and nanorotators. Any application which is based on photonics at mesoscales should take into account the new movements predicted here.
\end{abstract}

\pacs{}


\maketitle

\section{Introduction}

It has been known for more than a century that light carries linear and angular momentum \cite{Poynting1884,Loudon2012} in addition to energy. When light is scattered or absorbed by a particle, the transfer of momenta can cause the particle to move and/or to rotate. Thus light can be used to manipulate particles, molecules or mesoscale objects in general \cite{Grzegorczyk2006,Marago2013,Spesyvtseva2016,Gao2017}. Direct manipulation of objects through light-induced forces has led to formidable progress which has been impacting research in many areas ranging from ultra-cold matter physics \cite{Letokhov2007}, biology \cite{Ashkin1987,Svoboda1994}, microfluidics \cite{Rodrigues2017,Paie2018}, optical printing \cite{Gargiulo2017}, to optical engineering \cite{Li2008,Renaut2013,Qiu2014,Ma2012,Merklein2017} among other fields. For example, demonstration of levitation and trapping of micron-sized particles by radiation pressure dates back to 1970 \cite{Ashkin1970}. Since the 90's, even new states of matter have been conceived with manipulation by optical forces \cite{Burns1990}. In particular, one important consequence of electromagnetic two-particle interaction is the optical binding (OB) first noted by Burns et al. \cite{Burns1989,Figliozzi2017}. Two particles excited by a common field can form a bound dimer with stable or pseudo-stable positions \cite{Johnson1993,Figliozzi2017,Nan2018}.

Equally important to getting such manipulable forces is to realize proper theoretical/predictive models. In particular, the optical forces exerted on a small, single scatterer are typically assumed as the contributions due to its dipole moments \cite{Gordon1973}. In this regime, two force components are usually found, namely the gradient component and the scattering force \cite{Chaumet-Nieto2000}. Moreover, this last component is usually associated with the radiation pressure exerted by light on the system \cite{Novotny2012}. These components can be easily distinguished when the light is, for instance, a single plane wave. However, non-trivial contributions can exist for ``structured'' light having spin densities of light \cite{Mole2009}. This means that the more complex is the electromagnetic field around the system, the more complex are the exerted components of force and torque to model \cite{AndrewsBookStrLight,Nieto2015,Nieto2015b}. 

For coupled systems as dimers, there are no analytical formulations of the force far beyond the dipole-dipole coupling, even for simple illumination as with plane waves \cite{Kall2010,Albella2013,Bakker2015}. In the case of two bodies, it is commonly assumed that the system can undergo binding forces in addition to the scattering forces, both kinds generated by field gradients in response to the incident field \cite{Dholakia2010}. The former forces are contributions that try to attract/repel the particles each other and the latter forces, or the radiation pressure, push the system to the forward direction with respect to the direction of the incident wave. Moreover, the binding forces exerted on the system are expected to be compensated in symmetric or reciprocal systems, as it occurs for the case of a homodimer system, i.e. two coupled and equal spheres made of an isotropic material, under symmetric illumination. Many works have studied the optical forces exerted in dimers under the influence of light fields \cite{Chaumet2001,Sukhov2015,Kostina2017,Kall2010,Liaw2016}. In particular, Haefner et al. \cite{Haefner2009} have discovered the presence of spin torques in homodimers of nanospheres when illuminated symmetrically with plane waves having linear or circular polarization. However, many new degrees of freedom have been found in this work for the movement of electromagnetically coupled systems that have general validity irrespective of material and shape properties. A complete scheme of induced torques under linearly-polarized incident waves are reported here. Aditional spin contributions to the movement of the system are found in addition to the usual exerted forces and torques predicted in \cite{Haefner2009}. These induced, non-trivial torques are a product of the multiple interactions occurring in the system, there is no need of using complex helical fields to obtain angular momentum transfer \cite{Nieto2015,Sukhov2017}. The spin motors or ‘‘nanorotators’’ have been traditionally discussed based on optical traps created with circularly polarized light or vortex beams \cite{Dienerowitz2008,Jones2009}. 

A well-known method of discrete dipoles, namely the discrete dipole approximation (DDA), is used to perform our calculations \cite{DeSousa2016,Draine1988,Draine1994,Yurkin2007,AbrahamE2017}. For the sake of simplicity, we focus the study on the response by dimers of silica nanospheres under three configurations of variation of the illumination. With the aim to give a simple explanation of the phenomena, the results are compared with similar results obtained by simplified systems consisting in a few dipole moments. With this procedure, it is deduced how these new torques appear naturally as a consequence of the inhomogeneous inner fields that are induced on the system in reaction to the incident wave.

Similar results have been reported recently for two-dimensional systems of bound cylinders calculated with an integral formulation \cite{AbrahamE2016,AbrahamE2018_Ag,AbrahamE2018_Si}. As the present work uses another method of solving Maxwell equations and other dimensions, the results of this paper complete those previous studies with a generalization of the phenomena that is independent of the materials, dimensions, and geometries involved. 

This paper also shows that in general the retardation effects cannot be neglected in a scattering problem. Even when a few multipolar terms are not taken into account, the dynamics of the system under time-harmonic fields can be seriously affected. In this way, the results have a key role for the correct design of optical nanoscale devices and nanorotators, because they predict new dynamics of the systems as a first approximation of their movements. Although an exact electromagnetic method is used here, neither thermal nor Brownian forces are considered \cite{Albaladejo2011}. Also, no ``dynamic'' forces are calculated, i.e. forces that take into account initial velocities and accelerations of the wires \cite{Grzegorczyk2006dynamics}. Then, no complete dynamics is obtained for the system. Yet, the new dynamical features presented here are essential for the functionality and efficiency of optical small devices and they should be taken into account for future applications involving photonic forces.

\section{Method}

The DDA is a well-known method of resolution of electromagnetic scattering problems. It solves the problem by dividing the scatterer's domain into small subvolumes which respond to the electromagnetic field by induced dipole moments. Small non-magnetic particles or subvolumes develop only an electric dipole moment in response to the light's electric field. A complete version of the DDA can be found in Refs.~\cite{DeSousa2016,AbrahamE2017} for instance. Let's summarize the method we require here for our calculations involving isotropic and non-magnetic materials that respond to a time-harmonic field. In the absence of currents inside the object, the electric field is given by the solution of the volume-integral equation \cite{Novotny2012}
\begin{equation}
\label{eq-VIE}
\mathbf E(\mathbf r) = \mathbf E_0(\mathbf r) + k^2_0 \int_{V} \hat G(\mathbf r, \mathbf r^{\prime})
[\epsilon(\mathbf r) -1] \mathbf E(\mathbf r^{\prime}) d\mathbf r^{\prime} .
\end{equation}
Here, $\mathbf E_0(\mathbf r)$ is the electric field of the incident wave, $k_0 = \omega/c$ is the magnitude of the vacuum wave vector, $V$ is the volume of the object, and $\hat G(\mathbf r, \mathbf r^{\prime})$ is the vacuum dyadic Green tensor \cite{Novotny2012}.

In the DDA approach, the previous integral equation is solved by discretizing the volume $V$ as
$V= \sum^N_{n=1} V_n$, where $V_n$ is the volume of a homogeneous region where the electric
field is assumed to be constant. Thus, Eq.~(\ref{eq-VIE}) now reads
\begin{equation}
\label{eq-VIE-DDA}
\mathbf E(\mathbf r) = \mathbf E_0(\mathbf r) + k^2_0 \sum_n \hat G(\mathbf r, \mathbf r_n)
\left(\epsilon(\mathbf r) -1\right) \mathbf E(\mathbf r_n) V_n .
\end{equation}

Defining the dipole moments as
\begin{equation}
\mathbf p_n = \epsilon_0 V_n \left(\epsilon(\mathbf r) -1\right) \mathbf E(\mathbf r_n) ,
\end{equation}
we can rewrite Eq.~(\ref{eq-VIE-DDA}) as
\begin{equation}
\label{eq-VIE-DDA-p}
\mathbf E(\mathbf r) = \mathbf E_0(\mathbf r) + \frac{k^2_0}{\epsilon_0} \sum_n 
\bar{\hat G}(\mathbf r, \mathbf r_n) \mathbf p_n ,
\end{equation}
where
\begin{equation}
\bar{\hat G}(\mathbf r, \mathbf r_n) = \frac{1}{V_n} \int_{V_n} \hat G(\mathbf r, \mathbf r^{\prime})
d\mathbf r^{\prime} .
\end{equation}
It can be shown that \cite{DeSousa2016}
\begin{equation}
k^2_0 \bar{\hat G}(\mathbf r, \mathbf r_n) \approx
k^2_0 \hat G(\mathbf r, \mathbf r_n) \;\; \mbox{if} \; \mathbf r \notin V_n 
\end{equation}
and
\begin{eqnarray}
k^2_0 \bar{\hat G}(\mathbf r, \mathbf r_n) & \approx & 
-\hat L_n/V_n + i k^2_0 \mbox{Im} \{ \hat G(\mathbf r_n, \mathbf r_n) \} = \\
& &  -\hat L_n/V_n + i k^3_0/(6\pi) \hat 1 \;\; \mbox{if} \; \mathbf r \in V_n .
\end{eqnarray}
Here, $\hat L_n$ is the so-called electrostatic depolarization dyadic
that depends on the shape of the volume element $V_n$ \cite{Lakhtakia1992,Yaghjian1980}. For cubic volume 
elements, the depolarization tensor is diagonal: $\hat L_n = (1/3) \hat 1$.

Thus, we can now rewrite Eq.~(\ref{eq-VIE-DDA-p}) for the internal field, 
$\mathbf E_n \equiv \mathbf E(\mathbf r_n)$, as follows
\begin{eqnarray}
\label{eq-En}
\left[ \hat 1 + \left( \hat L_n - iV_n \frac{k^3_0}{6\pi} \right) [\hat \epsilon_n
- \hat 1 ] \right] \mathbf E_n & = &  \mathbf E_{0,n} + \nonumber \\
k^2_0 \sum_{m \neq n} \hat G_{nm} [\hat \epsilon(\mathbf r_m) - \hat 1] V_m 
\mathbf E_m , & &
\end{eqnarray}
where $\mathbf E_{0,n} \equiv \mathbf E_0(\mathbf r_n)$, $\hat \epsilon_n \equiv 
\hat \epsilon(\mathbf r_n)$ and $\hat G_{nm} \equiv \hat G(\mathbf r_n, \mathbf r_m)$.

The left-hand side of Eq.~(\ref{eq-En}) can be defined as the exciting field
$\mathbf{E}_{\rm exc}(\mathbf r_n)$, i.e., the field that excites the $n$-volume element.
Now, defining the polarizability of the $n$-volume element, as
\begin{equation}
\label{eq-alpha-sphere}
\alpha_n = \frac{\alpha_{0,n}}{1- ik^3_0 \alpha_{0,n}/(6\pi)},\,\,
\alpha_{0,n} = 3 V_n \left( \frac{\epsilon_n -1}{\epsilon_n + 2} \right).
\end{equation}
where $\alpha_{0,n}$ is the quasistatic polarizability, Eq.~(\ref{eq-En}) can be rewritten as 
a set of coupled dipole equations for the exciting fields at each element
\begin{equation}
\label{eq-Eexc}
\mathbf E_{{\rm exc},n} = \mathbf E_{0,n} + 
k^2_0 \sum^N_{m \neq n} \hat G_{nm} \alpha_m \mathbf E_{{\rm exc},m} .
\end{equation}
It is worth stressing that this DDA formulation includes automatically the so-called 
radiative corrections \cite{Sipe1974,Belov2003,Albaladejo2010}, which are related to the 
imaginary part of the Green tensor, and it is thus fully consistent with the optical 
theorem. On the other hand, from the solution of Eq.~(\ref{eq-Eexc}), which constitutes 
a set of 3$N$ coupled linear equations for the exciting fields, one can get the dipole
moments and the total internal fields as follows
\begin{eqnarray}
\label{eq-pn}
\mathbf p_n & = & \epsilon_0 \alpha_n \mathbf E_{{\rm exc},n} \\
\label{eq-En-pn}
\mathbf E_n & = & \frac{1}{\epsilon_0 (\epsilon_n - 1) V_n}\mathbf p_n .
\end{eqnarray}

From the knowledge of the dipole moments and the internal fields, one can easily compute the different cross sections (scattering, absorption, and extinction) of the object or the system in question. In particular, the extinction cross section, which we use below, can be obtained as follows. Assuming a plane-wave illumination, $\mathbf E_0(\mathbf r) = \mathbf E_0 e^{i\mathbf k_0 \cdot \mathbf r}$, the extinction cross section is given by \cite{DeSousa2016}
\begin{equation}
\label{eq-Cext}
C_{\rm ext} = \frac{k_0}{\epsilon_0 |\mathbf E_0|^2} \sum^N_{n=1} 
\mbox{Im} \left\{\mathbf{E}^{\ast}_0(\mathbf{r}_n)\mathbf p_n \right\} .
\end{equation}
To calculate the net time-average optical force exerted on a body or on a system, it occurs as with the calculation of the optical cross sections; we must use the proper dipole moments that represent each scatterer region in the frame of the DDA. We consider here that each body $b$ is composed by $N_b$ dipole moments and then the subindex $n$ run only over all the dipoles of the chosen body, i.e. $n=1,2,..., N_b$. Notice, on the other hand, that in the Eqns.~(\ref{eq-Eexc}) we use $N$ or the number of dipole moments in the whole system. This number accounts for the multiple interactions between all the dipole moments. In general, $N\neq N_b$ for several coupled bodies. In this way, the components of the net force exerted on the body $b$ in question, are represented by
\begin{eqnarray}
\label{eq-FcDDA}
F_{i,b}=\sum^{N_b}_{n=1}F_{n,i}
\end{eqnarray}
The $i$-component of this force, $F_{n,i}$, can be obtained from the time-averaged force on a particle within the Rayleigh approximation \cite{Chaumet-Nieto2000}. This is 
\begin{eqnarray}
\label{eq-FcDDA}
F_{n,i}=\frac{1}{2}Re\{\mathbf{p}^t_n[\partial_i\mathbf{E}^{*}(\mathbf{r},\omega)|_{\mathbf{r}=\mathbf{r}_n}\}
\end{eqnarray}
The derivatives of the total field $\partial_i\mathbf{E}(\mathbf{r},\omega)|_{\mathbf{r}=\mathbf{r}_n}$ at the dipoles's position $\mathbf{r}_n$ of body $b$ can be obtained from the DDA Eqns.~(\ref{eq-Eexc}), giving \cite{Chaumet2007}:
\begin{eqnarray}
\label{eq-FcDDA}
\partial_i\mathbf{E}(\mathbf{r},\omega))_{\mathbf{r}=\mathbf{r}_n}=\partial_i\mathbf{E}_0(\mathbf{r},\omega))_{\mathbf{r}=\mathbf{r}_n}+ \nonumber \\ 
+ \frac{k^2_0}{\epsilon_0}\sum^N_{m=1, m\neq n}(\partial_i\mathbf{G}(\mathbf{r},\mathbf{r}_m))_{\mathbf{r}=\mathbf{r}_n}\mathbf{p}_m]\}
\end{eqnarray}

The optical torques can also be calculated from the DDA method, as given in Ref.~\cite{Chaumet2009}:
\begin{equation}
\label{eq-TqDDA}
\mathbf{N}_b=\sum^{N_{b}}_{n=1} \mathbf{r}_{n} \times \mathbf{F}_{n} + \frac{1}{2}\sum^{N_{b}}_{n=1} Re[\mathbf{p}_{n} \times (\mathbf{p}_n/\alpha_0)^{*}]
\end{equation}
The first term of Eq.~(\ref{eq-TqDDA}) involves the so-called extrinsic torque and the second term is called the intrinsic torque. Their significance has been discussed in Refs.\cite{Nieto2015,Nieto2015b,Chaumet2009}, among others, but the main difference between them lies in the \textit{spatial} dependence of the first term. However, it is important to mention that $\mathbf{N}_b$ can represent both orbital or spin torque (do not confuse the torque $\mathbf{N}_b$ with the number of dipole moments of the body, $N_b$). The character of spin or orbital of the torque of Eq.~(\ref{eq-TqDDA}) will depend on the choice of the reference system to define the positions $\mathbf{r}_n$ of the dipole moments that compose the examined body. We then deal with spin torques when the associated positions are taken with respect to the center of each body. Otherwise, we will assume that the reference system is located at the center of mass of the whole system and then we will set orbital torques. The forces involved in the calculation of the extrinsic torque are the forces exerted on each dipole $n$ that composes the particle $b$. 

\section{Results}

\begin{figure}[!h]
\begin{centering}
\includegraphics[width=7cm,height=7cm,keepaspectratio]{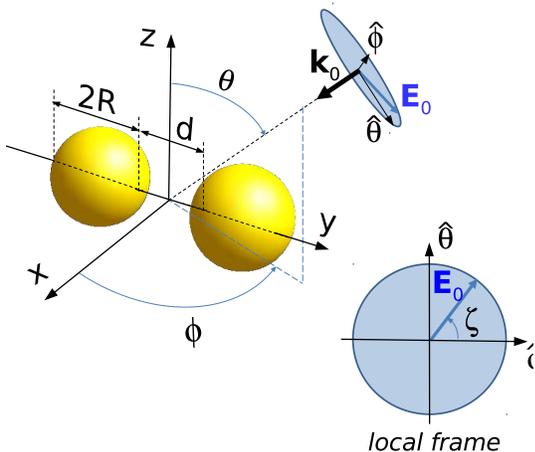}
\par\end{centering}
\caption{\label{fig:1_GralConfig}Scheme of the reference system and the geometrical configuration of the problem. The angles $\theta,\phi$ of the incident wave correspond to the spherical coordinate system. The inset graph shows the definition of the polarization angle $\zeta$. The electric-field vector is highlighted in blue.}
\end{figure}

The central configuration of this study can be seen in the Fig.~\ref{fig:1_GralConfig}. A homodimer of silica spheres is positioned with its axis parallel to the $y$-axis of a rectangular coordinate system. The gap is $d$ and the radii of the spheres are $R=240$ nm. The incident wave approaches with a wavevector $\mathbf{k}_0$ which is determined by the angles $\theta$, $\phi$ given by the spherical coordinates. The incident wave is fully determined by the incident wavelength $\lambda=2\pi/k_0$ and a third angle $\zeta$ which is the polarization angle -the electric field is highlighted in a blue letter, see also inset scheme-. Silica nanospheres are simulated by a refractive index $n=1.59 + i 10^{-7}$. \\

\textit{-Single Nanosphere} \\

\begin{figure}[!h]
\begin{centering}
\includegraphics[width=9cm,height=9cm,keepaspectratio]{figure2a.eps} 
\includegraphics[width=3cm,height=3cm,keepaspectratio]{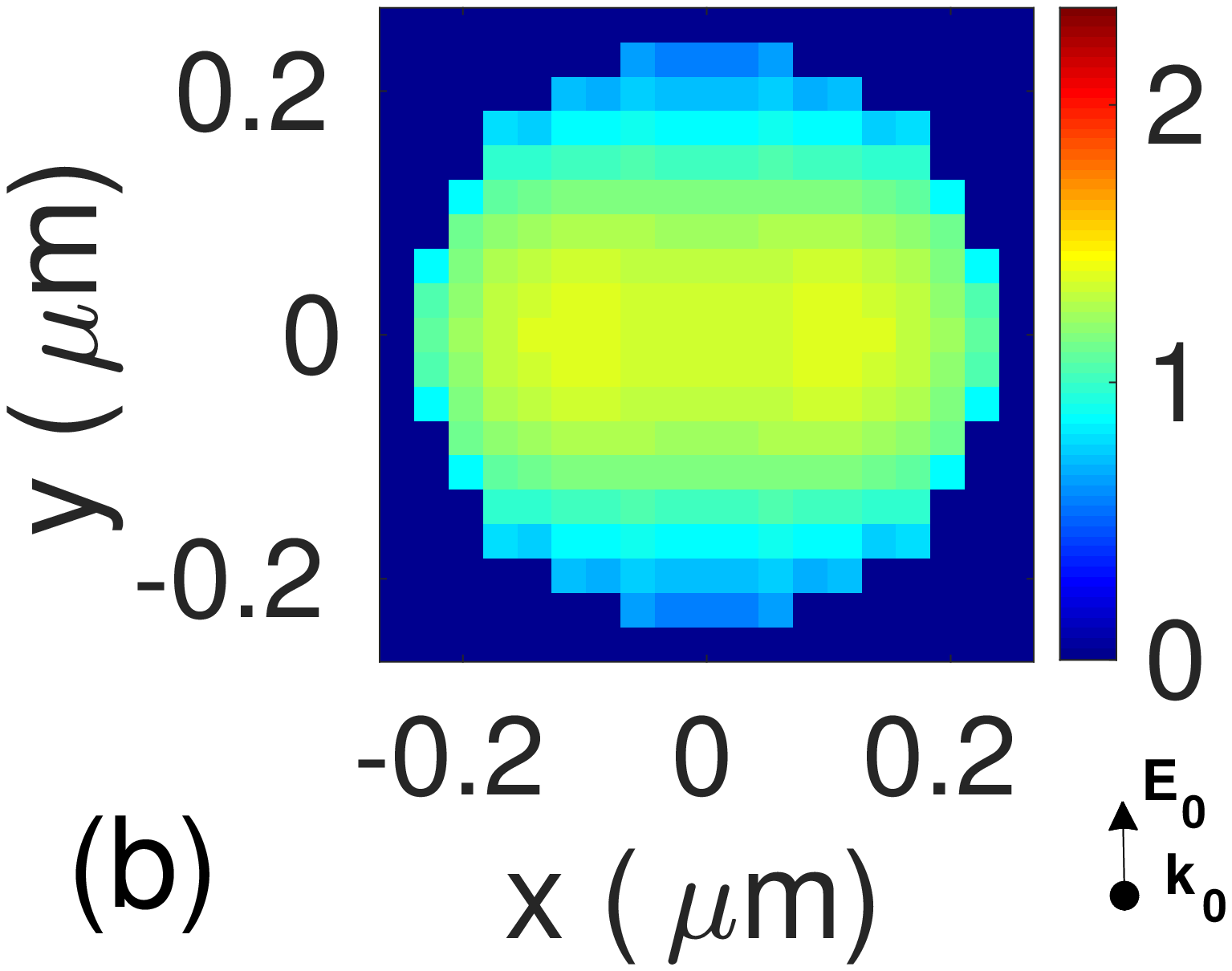}
\includegraphics[width=3cm,height=3cm,keepaspectratio]{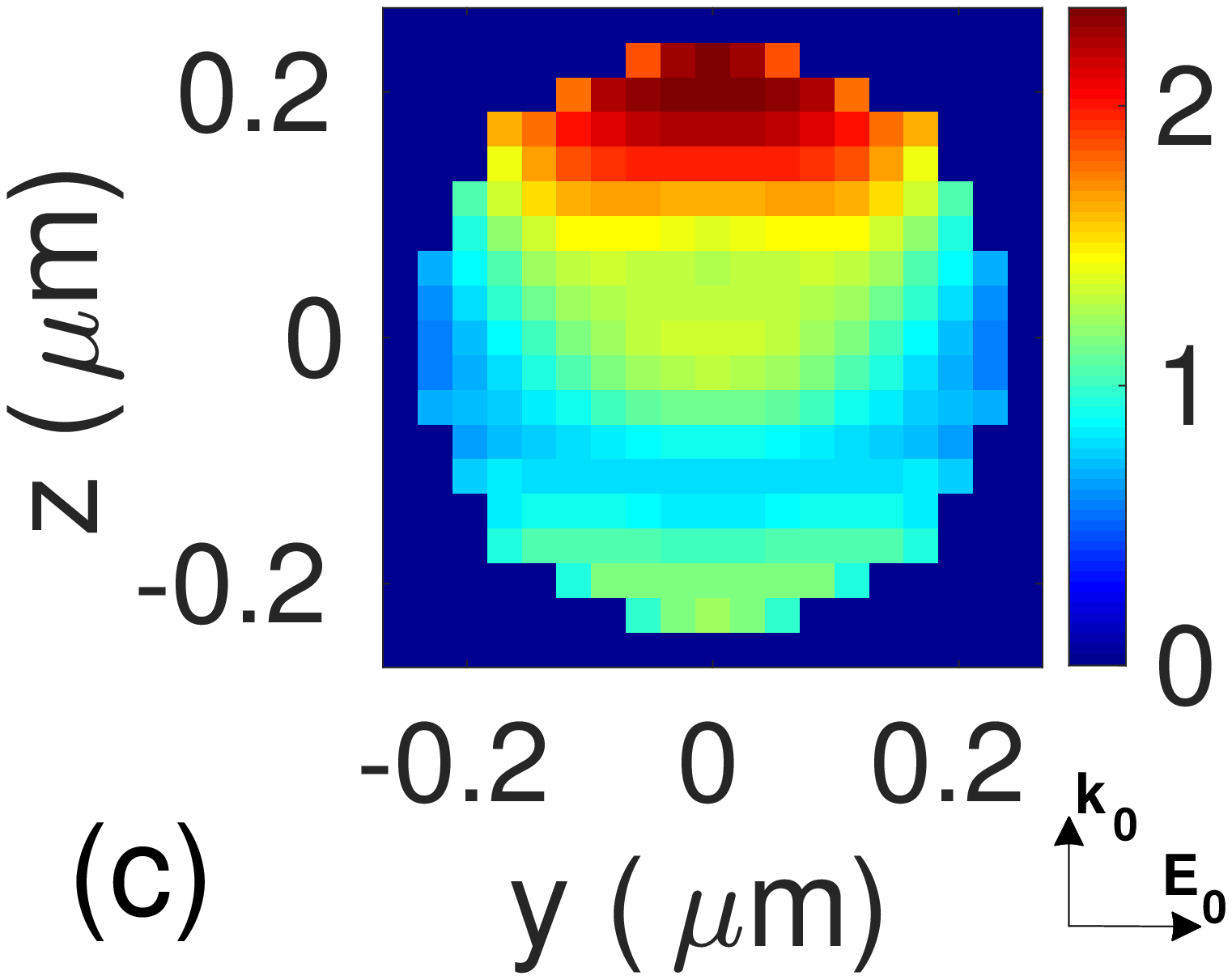}
\includegraphics[width=3cm,height=3cm,keepaspectratio]{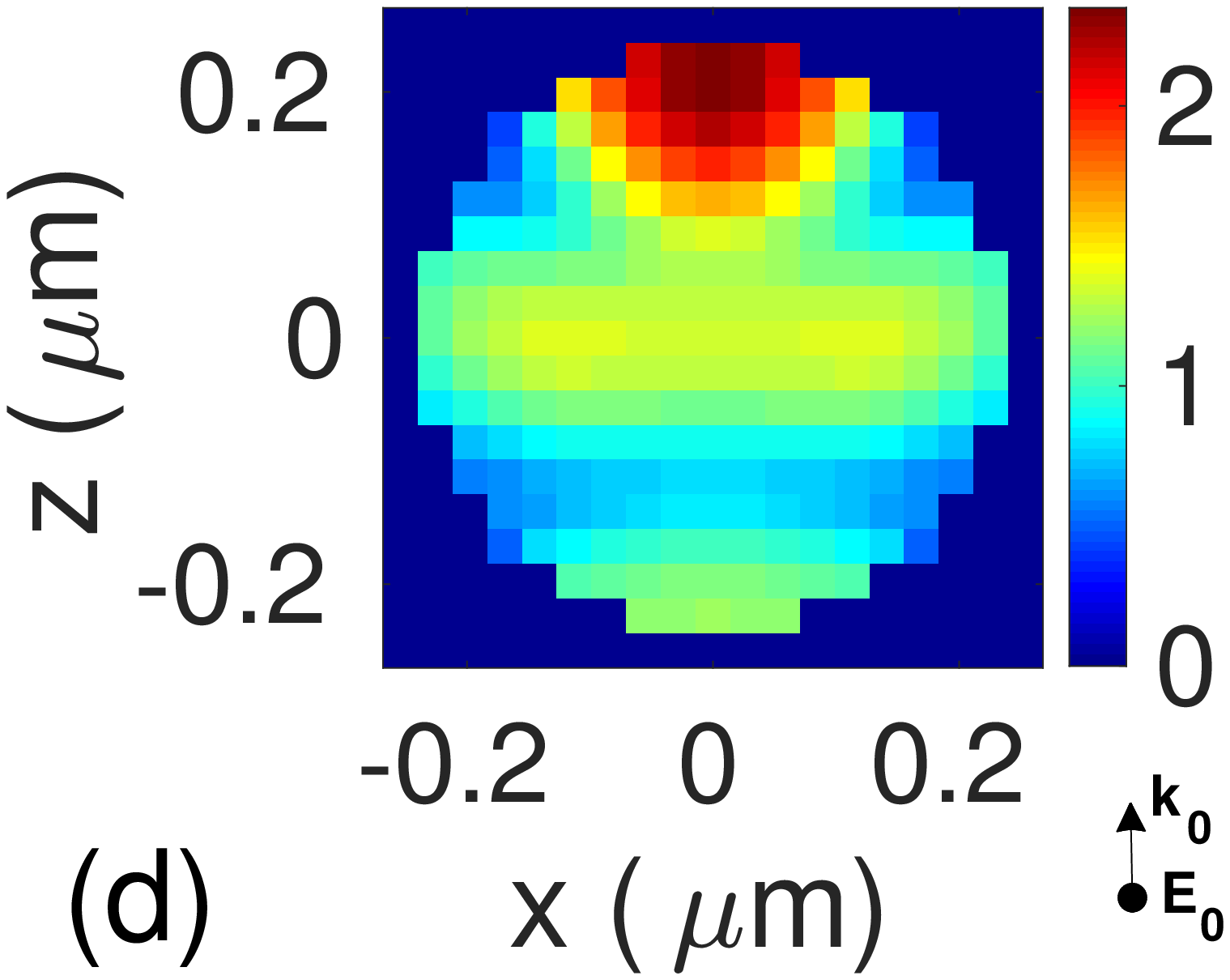} \\
\par\end{centering}
\caption{\label{fig:2_SingleSph}Morphology-dependent resonances in a silica nanosphere of $R=240$ nm. (a) Spectral curve for the extinction efficiency, $Q_{ext}$. (b)-(d) Maps of the relative module of the inner electric field, $|\mathbf{E}/\mathbf{E}_0|$,calculated at $\lambda=600$ nm with \#2553; (b) cut $x,y,z=0$, (c) cut $x,y=0,z$ and (d) cut $x=0,y,z$. The arrows indicate the direction of the vectors of the incident wave (not in scale).}
\end{figure}

In order to get a rapid insight into the spectral response of the silica nanosphere systems, the Fig.~\ref{fig:2_SingleSph} shows the optical response by a single silica nanosphere. For this study, the single sphere is located at the center of the coordinate system of the Fig.~\ref{fig:1_GralConfig}. In the extinction curve, we can see the multiple excitations due to the presence of morphology-dependent resonances (MDRs) \cite{VanBladel1977,Ng2005}. As the silica corresponds to a low refractive index, the resulting MDRs are overlapped in the optical region. At high energies, many modes appear in the spectrum and the overlapping is greater than at low energies. The spectrum was calculated with \#1791 dipole moments.

Although some changes are expected for the optical response of coupled silica nanospheres, the spectral structure may result quite similar in shape to the spectra of single spheres. This is somehow expected because the optical response by silica spheres is dominated by the presence of the MDRs which are volume resonances \cite{AbrahamE2018_Si}. To illustrate the volume nature of the MDRs, some inner field maps were calculated (see inset graphics) for an arbitrary value of wavelength, i.e. $\lambda=600$ nm. This particular value has been chosen because it corresponds to a spectral position where several overlapped modes can be excited, see the extinction behavior at this wavelength. The maps were calculated with \#2553 dipole moments.

Notice that the multiple scattering between all the dipoles originates an inhomogeneous inner field but this is always symmetric with respect to the incident wave field for a single sphere, see \ref{fig:2_SingleSph}(b-d). The resultant dipoles are, of course, oriented as the electric field or polarization vector (not shown) but they generate symmetric forces that give together the typical scattering component exerted on the sphere, i.e. the radiation pressure. This fact will be contrasted with the results obtained in the dimer system, giving interesting conclusions with regard to the movement of the system.

Here below, the optical response by dimers of silica spheres is explored in terms of the mechanical magnitudes. The results are scaled by proper factors, but they represent forces of an order of piconewtons when the intensity of the illumination reaches a few $mW/\mu m^2$. In the same way, the obtained torques reach an order of $pN.nm$ when the same order of illumination intensity is used.

\textit{-Dimer: Spectral variation.} \\

\begin{figure}[!h]
\begin{centering}
\includegraphics[width=9cm,height=9cm,keepaspectratio]{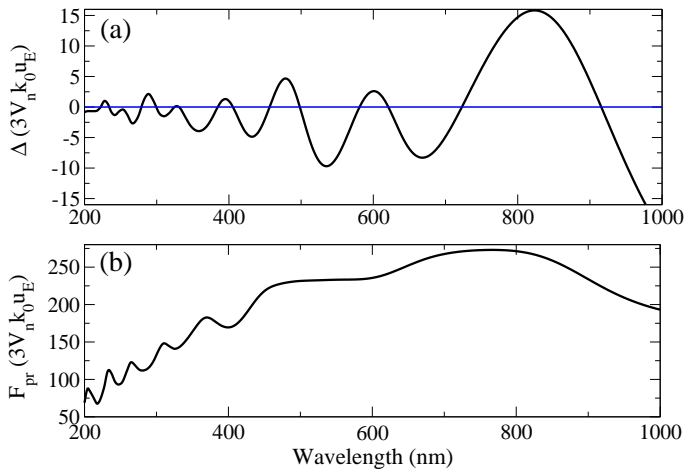}
\par\end{centering}
\caption{\label{fig:3_config1_DimerFcs}Spectra of scaled optical forces exerted on a silica dimer of $R=240$ nm and $d=500$ nm. The incident wave has a wavevector $\mathbf{k}_0=k_0\mathbf{\hat{z}}$ and is polarized along the dimer's principal axis, i.e. along with $\mathbf{\hat{y}}$. (a) Binding force, $\Delta=F_{1y}-F_{2y}$; the blue line indicates the zero of the scale. (b) Scattering force or radiation pressure, $F_{pr}=|F_{z}(CM)|$.}
\end{figure}

In order to study the spectral forces exerted on silica dimers, a first example has been taken for a gap of $d=500$ nm (Fig.~\ref{fig:3_config1_DimerFcs}). The incident wave is assumed to have a direction given by $\mathbf{k}_0=k_0\mathbf{\hat{z}}$ and a polarization $\mathbf{E}_0=E_0\mathbf{\hat{y}}$. The whole system suffers the action of binding forces, panel~\ref{fig:3_config1_DimerFcs}(a), due to the electromagnetic coupling between the spheres and the incident field while, at the same time, the system is also pushed by radiation pressure, panel~\ref{fig:3_config1_DimerFcs}(b), along with the forward direction. The binding force is defined as $\Delta=F_{1y} - F_{2y}$, the difference between the force components along the axis of the dimer. On the other hand, the scattering force is defined as $F_z=F_{z1} + F_{z2}$, the total force corresponding to the force exerted on the center of mass of the system. The forces are scaled to the magnitude $3V_nk_0u_E$, being $u_E=\frac{1}{2}\epsilon_0|\mathbf{E}_0|^2$ the electric energy density of the incident field and $V_n$ is the subvolume of the discretization used in the corresponding calculation.

The excitations of the MDRs in the dimer can also be seen through the spectra of Fig.~\ref{fig:3_config1_DimerFcs}. Notice how the binding force (panel~\ref{fig:3_config1_DimerFcs}(a)) alternates its sign by means of the excitations of the MDRs. Repulsive ($\Delta >0$) or attractive character ($\Delta <0$) is obtained depending on which resonance is excited: the curve oscillates around the zero-value of force (blue line). The values of the radiation pressure are much larger than those obtained for binding forces. In addition, some resonance shifts can be found between both kinds of spectra \cite{AbrahamE2018_Si,AbrahamE2018_Ag}. Thus, as seen in previous works, the optical forces of the system are able to ``feel'' the electromagnetic modes and they can even serve as near-field observables of the scatterer system.

\begin{figure}[!h]
\begin{centering}
\includegraphics[width=9cm,height=9cm,keepaspectratio]{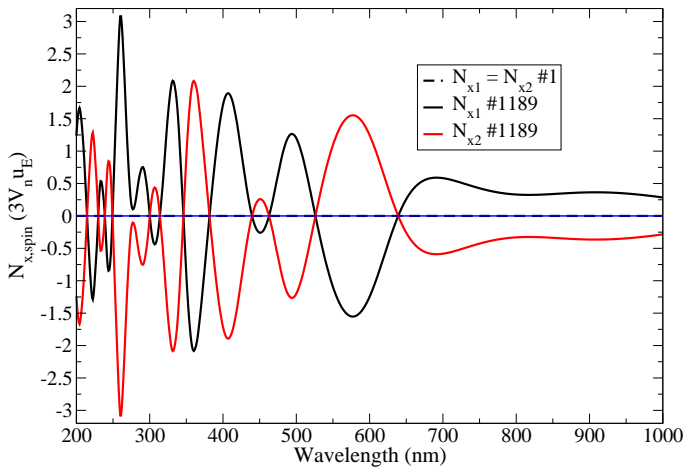}
\par\end{centering}
\caption{\label{fig:4_config1_DimerTqs}Spectra of scaled spin torques exerted on a silica dimer of $R=240$ nm and $d=500$ nm. The incident wave has a wavevector $\mathbf{k}_0=k_0\mathbf{\hat{z}}$ and it is polarized along the dimer's principal axis, i.e. along with $\mathbf{\hat{y}}$. The blue line indicates the zero of the scale.}
\end{figure}

For the configuration of illumination of the Fig.~\ref{fig:3_config1_DimerFcs}, neither spin nor orbital torque has been expected in the literature \cite{Haefner2009}. Logically, there are no orbital torques because we deal with a \textit{homodimer} \cite{AbrahamE2018_Si}, or a dimer made of similar particles. However, when introducing the DDA results, optical spin torques can be found with respect to the $x$-axis of the system (Fig.~\ref{fig:4_config1_DimerTqs}, see curves in solid line). This kind of spin has been recently found in 2D-systems by an integral formulation \cite{AbrahamE2016,AbrahamE2018_Ag,AbrahamE2018_Si}. The new dynamics can be obtained only when \textit{realistic} interactions between the scatterers are simulated, i.e. the complete multiple scattering must be taken into account. Specifically, the inhomogeneities induced in the inner fields are the cause of the spins. Notice that the phenomenon is exclusively obtained by bodies represented with multiple dipole moments; the induced spins for two bodies represented by single dipole moments are zero (Fig.~\ref{fig:4_config1_DimerTqs}, see dashed line). 

As it was said in previous works \cite{AbrahamE2016,AbrahamE2018_Ag,AbrahamE2018_Si}, the spin torques appear in \textit{coordinated} form when the dimer is composed by equal scatterers. In other words, the spin curves for each particle have equal values but opposite signs, see the curves in black solid line and in red solid line in Fig.~\ref{fig:4_config1_DimerTqs}. This means that the induced torques cancel each other and give zero torque for the whole system by reasons of symmetry.

\begin{figure}[!h]
\begin{centering}
\includegraphics[width=9cm,height=9cm,keepaspectratio]{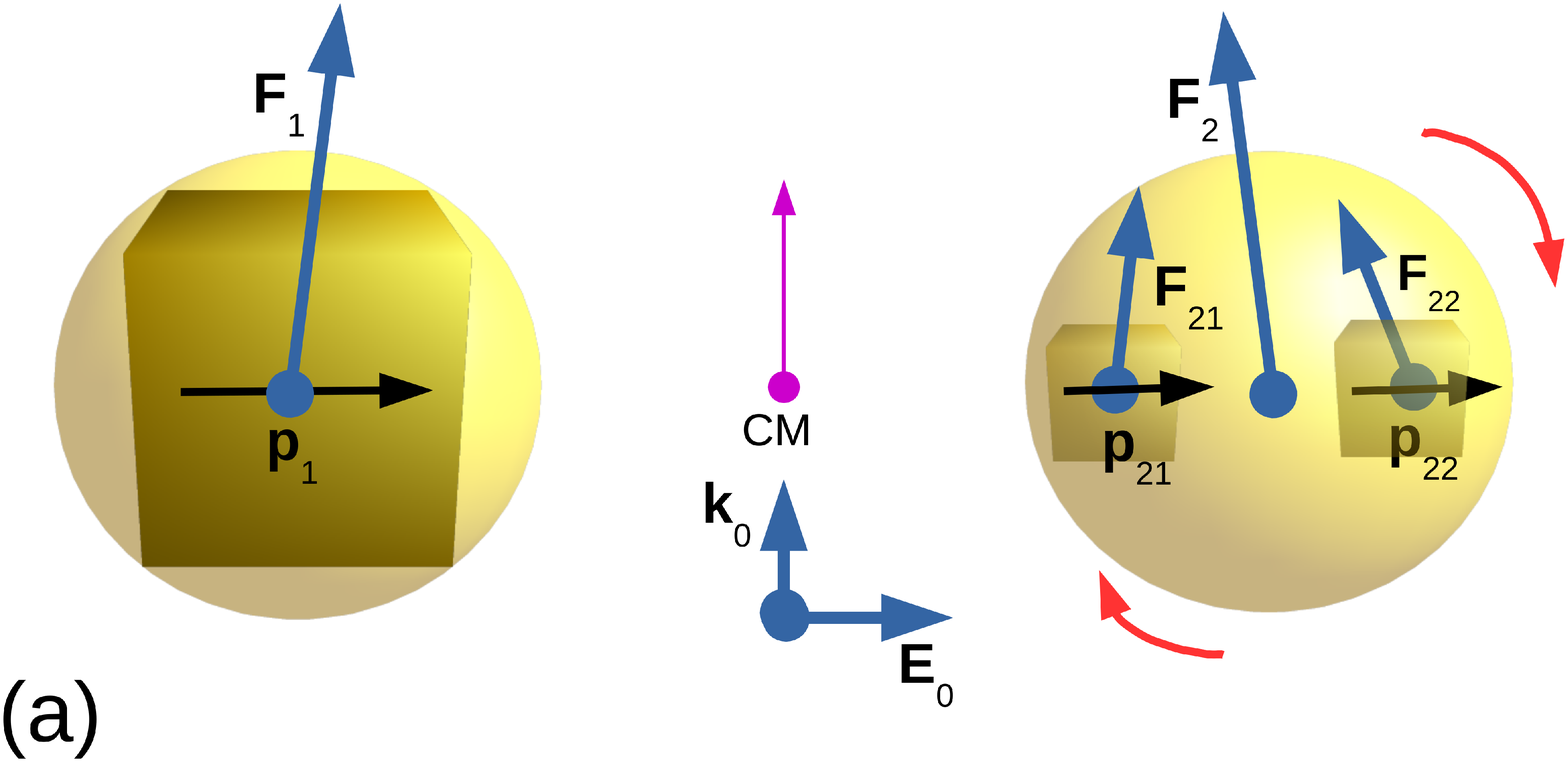} \\ 
\includegraphics[width=4cm,height=4cm,keepaspectratio]{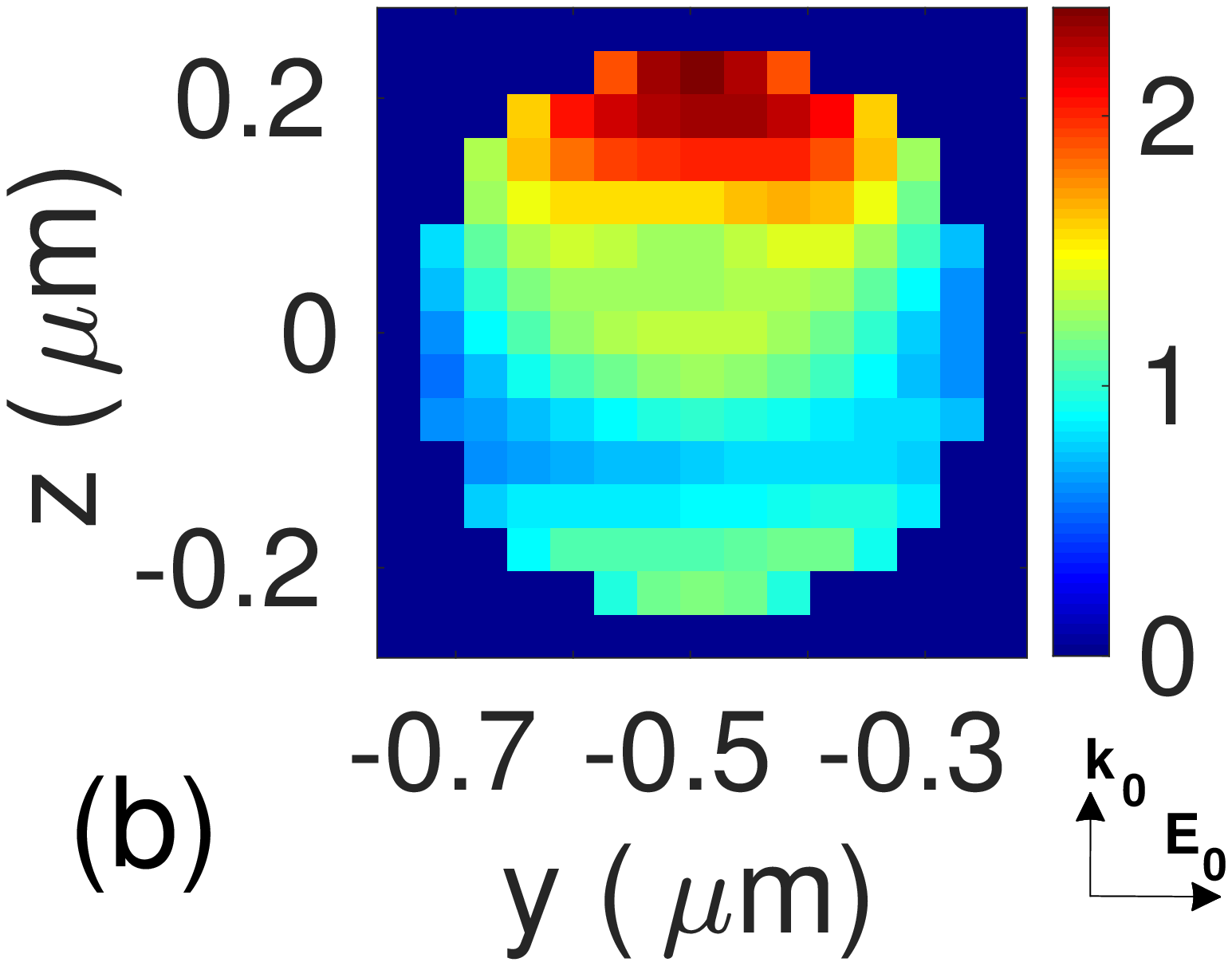}
\includegraphics[width=4cm,height=4cm,keepaspectratio]{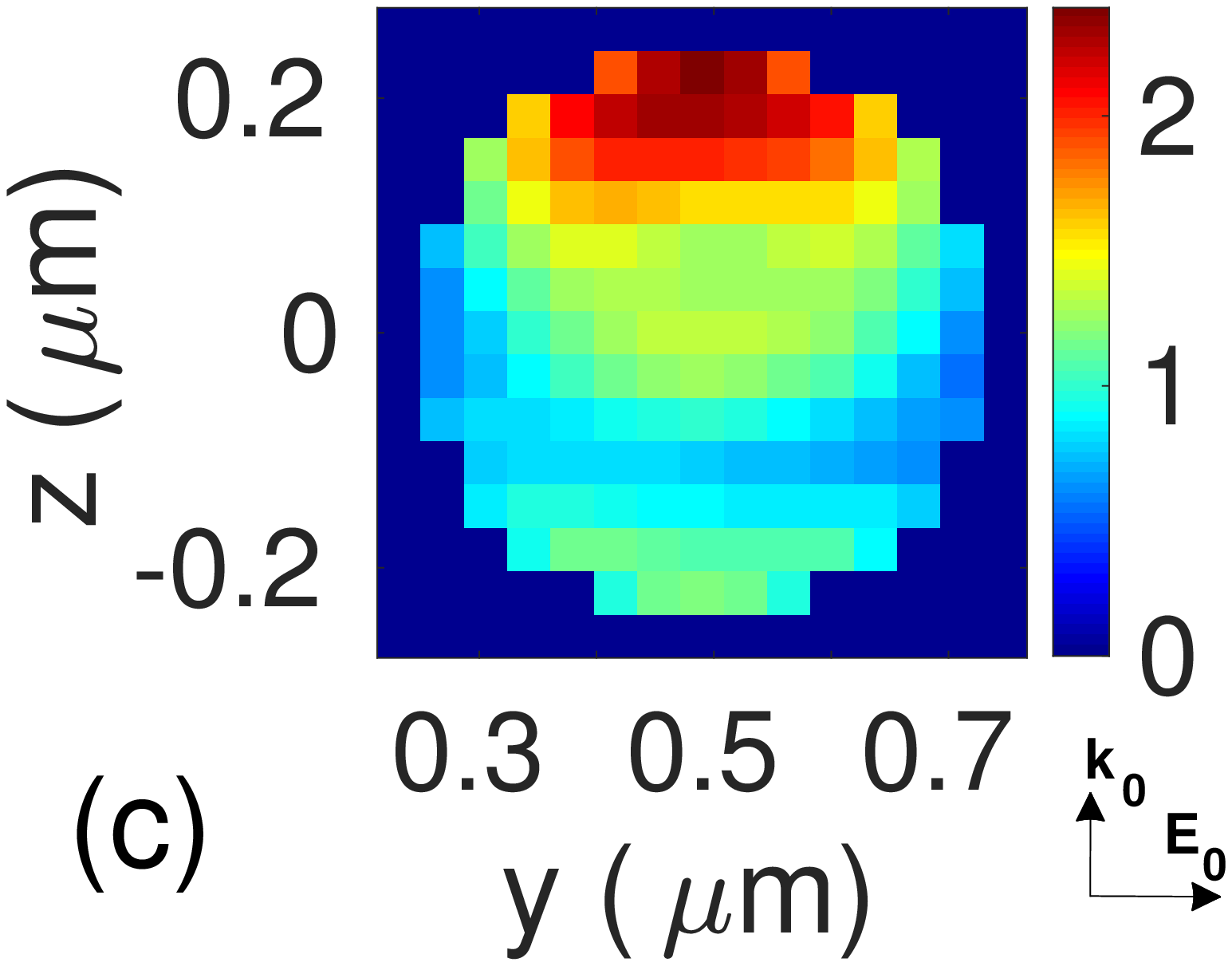}
\par\end{centering}
\caption{\label{fig:5_config1_3dipoles}(a) Scheme to explain the origin of the spin torques in a homodimer under the illumination configuration of the Figs.~\ref{fig:3_config1_DimerFcs} and \ref{fig:4_config1_DimerTqs}. The induced inner field in the particle $2$ make it spin around its center (red arrows) because the elemental forces $\mathbf{F}_{21}$ and $\mathbf{F}_{22}$ are unbalanced. The vector in magenta for the center of mass of the system (CM) represent the whole movement of the system but it is not scaled to the forces $\mathbf{F}_1$, $\mathbf{F}_2=\mathbf{F}_{21} + \mathbf{F}_{22}$ induced on each respective particle. The boxes represent the subvolumes of the dipole moments in the frame of the DDA method. (b)-(c) Maps of the relative module of the inner electric field, $|\mathbf{E}/\mathbf{E}_0|$  for the configuration of the Figs.~\ref{fig:3_config1_DimerFcs} and \ref{fig:4_config1_DimerTqs} at $\lambda=600$ nm (calculated with \#1189 dipole moments each sphere). The planes $y,z$ shown corresponds to the cut $x=0$. (b) Left sphere, $1$. (c) Right sphere, $2$.}
\end{figure}

We saw in the previous study for a single sphere that the particles must be represented by multiple dipole moments to obtain inhomogeneous fields. The simplest dimer configuration to obtain spins is a system represented by three dipole moments, see the panel (a) of Fig.~\ref{fig:5_config1_3dipoles}. In this case, one sphere is represented by a single dipole moment and the other sphere is represented by two dipole moments placed in symmetric positions with respect to the direction of polarization. In this way, notice that the spin can only be induced on the sphere composed of two dipole moments, see the red arrows on the scheme \ref{fig:5_config1_3dipoles}(a). The spin for the sphere $1$ in \ref{fig:5_config1_3dipoles}(a) is identically zero. The resultant inhomogeneity of the inner field inside sphere $2$ makes the whole particle to rotate because the induced tangential forces for each dipole's subvolume are different between them (inner field not shown in the figure for this configuration). However, the symmetry of the whole system is preserved from the point of view of the net forces. The induced force $\mathbf{F}_1$ on the sphere $1$ is perfectly balanced with the induced force $\mathbf{F}_2=\mathbf{F}_{21} + \mathbf{F}_{22}$ on the sphere $2$ along the polarization direction. As a result, a binding force (i.e. relative attraction/repulsion) between the particles can exist but the system as a whole moves forward due to the radiation pressure. Under this symmetric illumination, there is no unbalanced force along the direction given by $\mathbf{\hat{y}}$.

A similar configuration to the figures \ref{fig:2_SingleSph}-\ref{fig:4_config1_DimerTqs} has been studied through dimers of 2D-systems. The induced spin torques have also been thought as near-field observables of the interactions of the system. The spin torques have been shown to provide different information from the one that can be obtained by other mechanical or optical magnitudes. Notice that in general, the number of resonances ``detected'' has increased in the spectra of spins with respect to the number detected by the spectra of forces. In addition, the resolution of the resonances has improved, in general, in the spectra of torques. Another feature to be remarked upon is the shifts that exist in many spectral locations between the torque's and the force's resonances. Within all these features, however,  the results of the Fig.~\ref{fig:4_config1_DimerTqs} constitute a generalization in three dimensions of the results obtained by the previous studies referenced. As in those studies, the new phenomenon can be also obtained by plasmonic systems. In that case, the electromagnetic resonances detected correspond to surface resonances seen in the spectra as excitations of surface plasmons \cite{AbrahamE2016,AbrahamE2018_Ag}.

Although the spin torques are not familiar to us for this symmetric system, they appear as a natural phenomenon after the examination of the simpler case schematized in \ref{fig:5_config1_3dipoles}(a). Now we can observe the map of \ref{fig:5_config1_3dipoles}(b-c) where the distributions of the inner fields have been calculated for the configuration of the Figs.~\ref{fig:2_SingleSph}-\ref{fig:4_config1_DimerTqs} at the wavelength $\lambda=600$ nm. The panel \ref{fig:5_config1_3dipoles}(b) (\ref{fig:5_config1_3dipoles}(c)) corresponds to the cut $x=0$ of the field distribution $y,z$ inside the left (right) sphere, $1$ ($2$). We can see that the fields inside the spheres present almost the same distribution as in the map $y-z$ shown for the single sphere, Fig.~\ref{fig:2_SingleSph}. However, the distribution changes in this case due to the interaction between the spheres. Although we have a symmetric configuration with respect to the incidence, the multiple scattering between the spheres results in bent patterns for the inner fields. Consequently, those asymmetric patterns generate the unbalanced forces that produce the spin torques.

With the aim of exploring the new degrees of freedom that can be induced in dimers, other configurations of illumination have been studied. In the following, the results correspond to angular variations of the illumination at a fixed value of the wavelength, namely $\lambda=600$ nm. This value has been picked up in connection with the relative maximum obtained in Fig.~\ref{fig:4_config1_DimerTqs} at this spectral position. The results can be approximately reproduced in any experiment having any laser wavelength near this value as, for instance, an He-Ne laser $\lambda=632.8$ nm.  \\

\begin{figure}[!h]
\includegraphics[width=9cm,height=9cm,keepaspectratio]{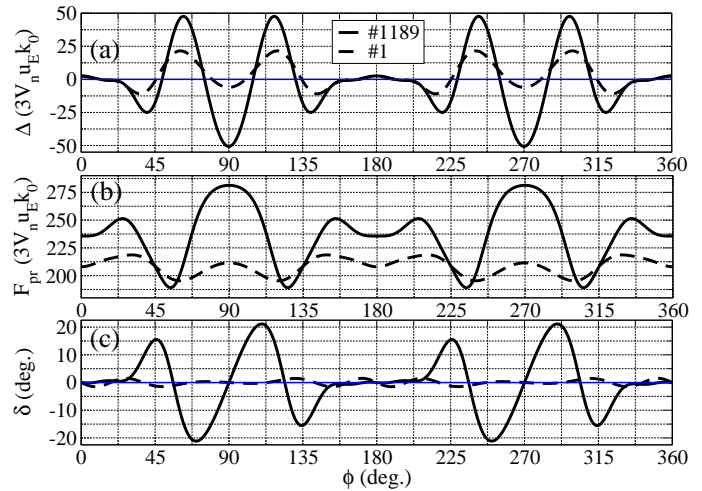}
\caption{\label{fig:6_config2_DimerFcs}Scaled optical forces exerted on a silica dimer of $R=240$ nm and $d=500$ nm when the angle $\phi$ of the illumination is varied. The incident wavelength is $\lambda=600$ nm, and the other angles are $\theta=90$ deg. and $\zeta=0$ deg. (a) Binding force. (b) Module of the radiation pressure for the whole system, $F_{pr}=|\mathbf{F}_{pr}|=\sqrt{F^{2}_{x,CM}+F^{2}_{y,CM}}$. (c) Deviation $\delta=\phi(\mathbf{F}_{pr})-\phi$ of the radiation pressure from the direction of the incident wave. Curves in continuous (dashed) line correspond to a calculation with $\#1189$ ($\#1$) dipole moments per particle. The blue lines in (a) and (c) indicate the zero of the scales.}
\end{figure}

\textit{-Dimer: Azimuthal variation.} \\

The variations of the induced forces with the azimuthal angle can be seen in the Fig.~\ref{fig:6_config2_DimerFcs}. The panels show the binding forces (a), the module $F_{pr}=|\mathbf{F}_{pr}|=\sqrt{F^{2}_{x,CM}+F^{2}_{y,CM}}$ of the scattering force (b) and (c) the deviation of direction of the scattering force with respect to the direction of the incident wave, $\delta=\phi(\mathbf{F}_{pr})-\phi$. Here the angle of the scattering force is defined as $\phi(\mathbf{F}_{pr})=atan(F_{y,CM}/F_{x,CM})$ and the components of the center of mass are defined as $F_{j,CM}=F_{j,1} + F_{j,2}$ where $j=x,y$. The curves in solid line are calculated with $\#1189$ dipole moments per particle and they are compared against the same results but simulating the particles by single dipole moments (dashed line, $\#1$ each). The other angles of the illumination are set at $\theta=90$ deg. and $\zeta=0$ deg.

All the curves are periodic with the variation of $\phi$. Moreover, these curves present a modulation induced by the multiple scattering at $\lambda=600$ nm. We can find angular shifts between common features obtained for each curve. The binding force presents a small value of attraction at $\phi=0,180$ deg (see the curve in solid line) and it changes as $\phi$ is increased, oscillating from attraction to repulsion several times. The absolute minima of the curve are reached at $\phi=90,270$ deg. The minima correspond to the strongest states of repulsion of the configuration because the phase difference of the incident field between the particles is the greatest at these angles of illumination. In other words, the end-fire illumination is the most antisymmetric than it can be obtained from the point of view of the dimer. On the contrary, the values of the radiation pressure in \ref{fig:6_config2_DimerFcs}(b) have the absolute maxima at these right angles.

Notice that the force values of \ref{fig:6_config2_DimerFcs}(a) and (b) for $\phi=0$ deg. correspond exactly to their respective analogues in the panels \ref{fig:3_config1_DimerFcs}(a) and (b) for the wavelength $\lambda=600$ nm. In particular, this is useful to compare the values obtained in \ref{fig:6_config2_DimerFcs}(b) against the constant value for a single sphere, which can be obtained by calculating the radiation pressure exerted on the sphere $R=240$ nm at any arbitrary azimuthal angle. This value is approximately half of the value of $F_{pr}$ for the system at $\phi=0$ deg., i.e. $F_{pr,b=1}\approx 118$ in the units $3V_nk_0u_E$ of the scaled forces. Thus, the analysis of the variations of $F_{pr}$ with the azimuthal angle also serves to estimate the effects of the realistic interaction between the spheres if compared with the single-sphere response.

Logically, the contributions to the force values are changed by the number of dipole moments that represent the particles as we can see from comparing the curves in solid vs. in dashed line. However, this difference manifests also in a new phenomenon that has never been previously discussed in the literature. This is, the angle of the net scattering force for the system may result quite different than the angle of the illumination. Moreover, this angle depends on the inner-field modulation of the MDRs excited. As a result, the whole dimer moves as more on the right or more on the left than expected by the forward direction. Thus, the deviation of \ref{fig:6_config2_DimerFcs}(c) depends on the relative phase modulation of the field that is set by the angle of illumination, $\phi$. In the current example of the dimer, we can observe from \ref{fig:6_config2_DimerFcs}(c) that $\delta$ reaches extremals of $\approx21$ deg. at the angles $\phi \approx70,110,250,290$ deg and of $\approx15.5$ deg. at the angles $\phi \approx45,135,225,315$ deg.

The wide variation of deviation is an exclusive phenomenon induced by the multipolar structure of the scattering of the system; notice that there is also an angular deviation for the interaction between the single dipole moments (case $\#1$, see curve in dashed line in Fig.~\ref{fig:6_config2_DimerFcs}(c)). In this case, the whole system can be seen as composed by two dipole moments for the case $\#1$, and then the response of this system is quadrupolar in general \cite{Jackson}. If the illumination is not aligned with any principal direction for this quadrupole, the radiation pressure results deviated. However, the effect of deviation is vanishing for this case; the maxima do not exceed the $1.6$ \%. In the case of single dipole moments, the inner fields have constant values but they vary smoothly by the change of the relative phases that influences the dipoles' interaction. In this way, it could be practically considered that the angle of the scattering force coincides with the angle of illumination. Interestingly, the effect preserves the symmetry of the configuration; for the angles $\phi=0,90,180,270$ deg there is no induced deviation of the scattering angle (see both curves in solid and in dashed line). In the case of a single sphere, there is no deviation from the forward direction because the inner field responds only to the incident field, no matter how many dipole moments compose the sphere. The induced field results always symmetric with respect to the illumination direction and makes the particle to follow the forward direction, no matter how inhomogeneous the field is.

\begin{figure}[!h]
\begin{centering}
\includegraphics[width=9cm,height=9cm,keepaspectratio]{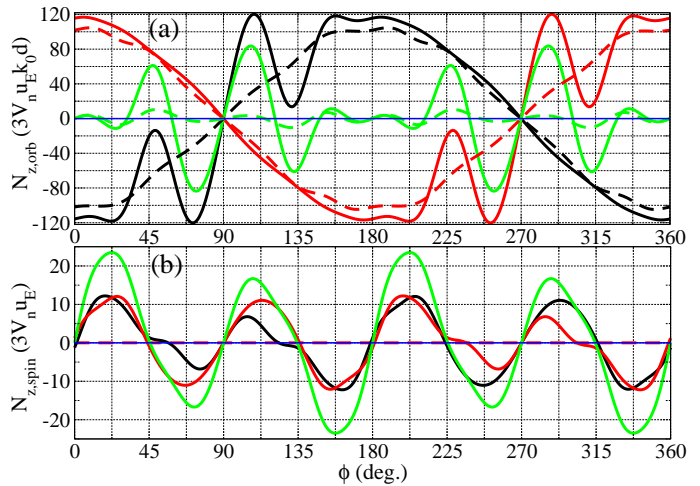}
\par\end{centering}
\caption{\label{fig:7_config2_DimerTqs}Scaled optical torques exerted on a silica dimer of $R=240$ nm and $d=500$ nm when the angle $\phi$ of the illumination is varied. The incident wavelength is $\lambda=600$ nm, and the other angles are $\theta=90$ deg. and $\zeta=0$ deg. (a) Orbital torque; (b) Spin torques. Curves in continuous (dashed) line correspond to a calculation with $\#1189$ ($\#1$) dipole moments per particle. Black (red) line for torque $N_{z1}$ ($N_{z2}$) exerted on the sphere 1 (2), green line for torques $N_{z}(CM)=N_{z1}+N_{z2}$ exerted on the center of mass of the system.}
\end{figure}

Now we explore more degrees of freedom to the study of the dynamics of the system. The investigation of the torques exerted under this configuration is presented in the Fig.~\ref{fig:7_config2_DimerTqs}. The results for both cases $\#1189$ and $\#1$ have been also added, see curves in solid and in dashed line. In this case, the symmetry breaking induced by the illumination direction with respect to the dimer axis allows for the presence of orbital torques in the structure (panel \ref{fig:7_config2_DimerTqs}(a)) even in the case $\#1$. In agreement with the broken symmetry, the spin torques exerted on each particle result different in general (panel \ref{fig:7_config2_DimerTqs}(b)). The only non-zero components correspond to torques aligned with the $z$-axis. The torques are influenced by a modulation similar to the one found for the forces in Fig.~\ref{fig:6_config2_DimerFcs}. Similar features are repeated with the period of $180$ deg. and again, the results under right angles of illumination save the symmetry of the configuration. In order to approach the movement for the whole system, the curves in green line have been added to the figures. They represent the resultant torques for the center of mass of the system, as given by $N_z(CM)=N_{z1}+N_{z2}$ for both the orbital \ref{fig:7_config2_DimerTqs}(a) and the spin components \ref{fig:7_config2_DimerTqs}(b). Notice that all the torques for the system vanish when the illumination direction is given by the values $\phi=0,90,180,270$ deg. In particular, the values of the spin torques for $\phi=0$ deg and $\lambda=600$ nm of Fig.~\ref{fig:4_config1_DimerTqs} are recovered in the panel~\ref{fig:7_config2_DimerTqs}(b) if a transformation of the axis of rotation $x \rightarrow z$ is performed. 

Again, the spin torques result in zero for the case of two dipole moments, see dashed line in \ref{fig:7_config2_DimerTqs}(b). On the other hand, the orbital torques present quasi-sinusoidal curves for each particle in the case \#1, see curves in black and in red line in \ref{fig:7_config2_DimerTqs}(a); there is a modulation for the curves of $\#1$ but it is much weaker than the one appearing in the curves for $\#1189$. In the case of the spheres with $\#1189$, the orbital components for each scatterer are ruled by this modulation that is mounted over the sinusoidal function. Thus, the orbital torques can be interpreted as a result of a combination of dipole-dipole interactions with multipolar contributions added to them. In this way, the predicted movement is not trivial; the spin components also results in oscillating curves. Both particles follow similar phases in a wide range of values of $\phi$ so they would spin in the same sense of rotation in those angular values, giving a net spin for the whole dimer that vanishes at $\phi=\kappa 45$ deg (curve in green line in \ref{fig:7_config2_DimerTqs}(b)), being $\kappa$ an integer number.

Given the complex movement that the curves of the Figs.~\ref{fig:6_config2_DimerFcs} and \ref{fig:7_config2_DimerTqs} imply, we could obtain a rough approach to the movement of the system by taking only the curves $\#1$ as many works do throughout the literature \cite{Mohanty2004,Tumkur2016,Tumkur2018,Gargiulo2017}. The dipole-dipole interaction may result in a more intuitive interpretation of the electromagnetic coupling of the scatterers. This simple picture can explain some features of the interaction as the sign of the binding forces in \ref{fig:6_config2_DimerFcs}(b) (curves in dashed line) because the attraction/repulsion is seen as generated by the forces exerted between equal or opposite charges in the dipole moments. It can also explain the origin of the orbital torques in \ref{fig:7_config2_DimerTqs}(a) for the case \#1 as a result of the alignment of the total dipole of the system with the field. The picture of the dynamics of the system based on the alignment of dipole moments should be managed with care because the induced dipole moments have complex components. Two perpendicular vectors of the dipole moments can be defined with respect to the incident field, namely the parallel and perpendicular components to it. They are both complex quantities and different from zero under illumination with an arbitrary angle $\phi$. Yet, the particular situation under illumination with right angles, i.e. $\phi=0,90,180,360$ deg., results different because the perpendicular components of the dipole moments are identically zero. Moreover, if a small particle approximation can be applied (of course it is not the general case), we can always get almost zero perpendicular components of the dipole moments no matter the value $\phi$ is and the imaginary parts of the induced dipole moments are compensated (compensation of the phases). It is in this context when the two-dipoles interpretation is useful. Still, if the system is observed under angles of illumination close to the right angles, we can see that the system tries to get one of two stable configurations \cite{Dholakia2010}, depending on the relative phases for each dipole; one is the position of canceled dipole moment for the system. The other one is the complete alignment of parallel dipoles to form a net dipole moment for the system. However, the phases of the dipole moments (imaginary parts of the components) are not compensated if the particles are not relatively small enough with respect to the wavelength, and this makes the dipole-dipole interpretation more difficult to deal with. Even more, the spin torques are completely avoided within the frame of this formulation. This is maybe the reason why the spins have not been predicted long time ago \cite{Haefner2009}. 

Then, in conclusion, the complex dynamics predicted here cannot be simply approached by taking the interaction between a few dipole moments in the system. The complete multiple scattering must be taken into account to observe all the induced mechanical features. \\

\textit{-Dimer: Variation of polarization.} \\

Due to the symmetry of the homodimer system, the study of the variation of the angle $\theta$ has no relevance and that configuration can be described by the rest of the analysis carried out up to here. In this subsection, the angle of polarization $\zeta$ is varied for the fixed wavelength $\lambda=600$ nm and a gap of $d=2R=480$ nm. In particular, this configuration of illumination has been studied previously in Ref.~\cite{Haefner2009} for silica dimers. However, some of the induced torques were not predicted; in this work, we complete the information about the movement of the system under this configuration. In the following, we set the other illumination angles at the values $\theta=180$ deg. and $\phi=0$ deg.

\begin{figure}[!h]
\begin{centering}
\includegraphics[width=9cm,height=9cm,keepaspectratio]{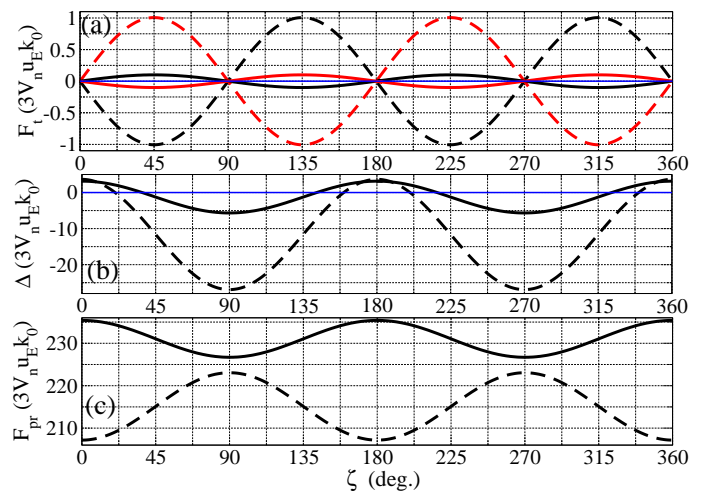}
\par\end{centering}
\caption{\label{fig:8_config3_DimerFcs} Scaled optical forces exerted on a silica dimer of $R=240$ nm and $d=480$ nm when the angle of polarization $\zeta$ of the illumination is varied. The incident wavelength is $\lambda=600$ nm, and the other angles are $\theta=180$ deg. and $\phi=0$ deg. (a) Tangential force $F_{t}\equiv F_{x}$, curves in black (red) line for force exerted on the sphere 1 (2); (b) Binding force; (c) Module of the radiation pressure. Curves in continuous (dashed) line correspond to a calculation with $\#1189$ ($\#1$) dipole moments per particle.}
\end{figure}

\begin{figure}[!h]
\begin{centering}
\includegraphics[width=9cm,height=9cm,keepaspectratio]{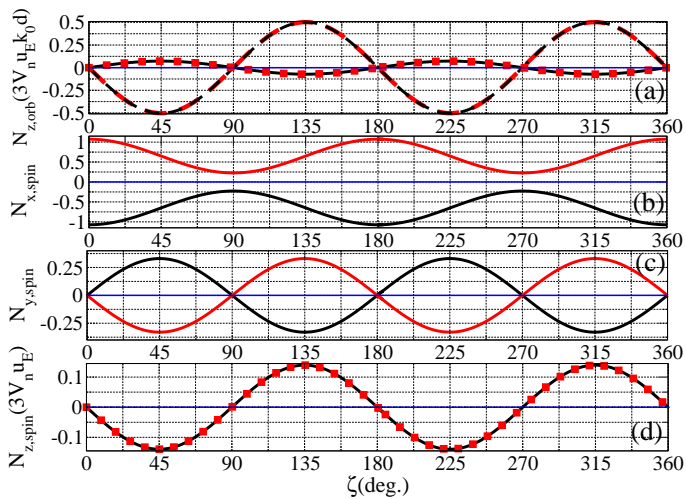}
\par\end{centering}
\caption{\label{fig:9_config3_DimerTqs}Scaled optical torques exerted on a silica dimer of $R=240$ nm and $d=480$ nm when the angle of polarization $\zeta$ of the illumination is varied. The incident wavelength is $\lambda=600$ nm, and the other angles are $\theta=180$ deg. and $\phi=0$ deg. and $\zeta=0$ deg. (a) Orbital torque; (b)-(d) Spin torques.  Curves in a continuous line or line with symbol correspond to a calculation with $\#1189$ dipole moments per particle. Curve in dashed line for $\#1$. Black (red) line for torques $N_{z1}$ ($N_{z2}$) exerted on the sphere 1 (2).}
\end{figure}

As we did before, we first analyze the behavior of the induced forces, Fig.~(\ref{fig:8_config3_DimerFcs}), for the cases of spheres made with \#1189 (curves in solid line) and \#1 (curves in dashed line). The curves in black (red) line in \ref{fig:8_config3_DimerFcs}(a) correspond to the net tangential force induced on the left (right) sphere, i.e. sphere $1$ ($2$). As pointed out in Ref.~\cite{Haefner2009}, the tangential forces \ref{fig:8_config3_DimerFcs}(a) may result unusual and already predict the presence of non-trivial orbital torques, see \ref{fig:9_config3_DimerTqs}(a). The Fig.~\ref{fig:9_config3_DimerTqs} follows the same color code as the Fig.~\ref{fig:8_config3_DimerFcs} with the exception of the curve in red solid line in panels (a) and (d) that is replaced for a curve with red symbols for clarity. The resultant binding forces are plotted in \ref{fig:8_config3_DimerFcs}(b) and the module of the radiation pressure is plotted in \ref{fig:8_config3_DimerFcs}(c). In this configuration, the direction corresponding to the radiation pressure is always directed parallel to the versor $\mathbf{\hat{z}}$, but its module varies smoothly with the polarization angle. Observe that all the curves present a clear sinusoidal form due to the phase variation of the incident field with respect to the symmetry of the dimer.   

The presence of the tangential forces is not a phenomenon exclusively occurring with many dipole moments (curves in solid line). The tangential forces are also induced in the case \#1 for each sphere (curves in dashed line). This means that there will be orbital torques also for \#1 when the angle $\zeta$ is varied, in a manner similar to the case when $\phi$ were varied. But more importantly, the effect of adding dipole moments for each sphere results in a very different dynamics in this particular case. Notice that the resultant tangential forces for the case \#1189 are reduced in absolute value and even change its sign with respect to the case of the two dipoles moments under this wavelength, see \ref{fig:8_config3_DimerFcs}(a). Such behavior is another illustration of the errors one can get when trying to describe the dynamics of the system with only a few dipole moments.

The variation of the binding force and the radiation pressure with the angle $\zeta$ are also reduced with the presence of several dipole moments, \ref{fig:8_config3_DimerFcs}(b) and (c). We could say that the multiple interactions between all the dipole moments moderates the variations of the forces exerted on the system. In particular, it can be seen from \ref{fig:8_config3_DimerFcs}(b) that the repulsion between the spheres is very attenuated with the multiple dipoles' response; the negative values are not so pronounced for \#1189 if compared to the case \#1. In addition, the curves for the tangential forces in \ref{fig:8_config3_DimerFcs}(a) differ in phase in $90$ deg. with respect to the curves of the forces in \ref{fig:8_config3_DimerFcs}(b) and (c). In other words, symmetric inductions as the radiation pressure and the binding force prevail for symmetric electric fields as with $\zeta=90,270$ deg. On the other hand, asymmetric inductions such as the tangential forces cancel out for these values of $\zeta$. 

Although not being realistic, it is interesting to describe the dipole-dipole dynamics of the dimer, see the curves in dashed line of the Figs.~\ref{fig:8_config3_DimerFcs} and \ref{fig:9_config3_DimerTqs}(a). Its utility was discussed in the previous subsection. Notice that there are two sets of angular positions where the system gains some stability. One set corresponds to $\zeta=0,180$ deg. where the system finds an equilibrium of zero tangential forces. However, it is unstable because a little angular perturbation around these positions would force the system to move off from its alignment with the incident electric field. Observe the values and signs of the tangential forces in \ref{fig:8_config3_DimerFcs}(a) and the torques in \ref{fig:9_config3_DimerTqs}(a). At $\zeta=0,180$ deg. the two complex dipole moments become parallel to the incident electric field in a way they are oriented along their connecting line ($\rightarrow \rightarrow$). Thus we obtain attractive forces between them because the opposite charges of the dipoles attract them to each other. However, this alignment is not preferred as soon as the system leaves the configuration of $\zeta=0,180$ deg.

The other equilibrium set is obtained for $\zeta=90,270$ deg. In this case, the system finds a stable equilibrium; the tangential forces are zero but small deviations from those angle values result in a ``rapid'' realignment to the original position. The system prefers to align its axis to the perpendicular direction with respect to the incident electric field. The two complex dipole moments become parallel to each other and to the incident electric field. Thus we obtain repulsive forces between them because the positive and negative charges of the dipoles coincide in position and repel each other ($\uparrow \uparrow$). Also note that for this set of angles, the values of the radiation pressure are greater and in opposite phase with respect to the binding force, see \#1 of \ref{fig:8_config3_DimerFcs}(b-c).

The dynamics for the case \#1189 is completely changed with respect to the dynamics for \#1. Although the signs of $\Delta$ do not change near $\zeta=0,90,180,270$ deg., the values of $F_t$ and $F_{pr}$ result very changed. Actually, $F_t$ results inverted in sign for all $\zeta$; \#1189 and \#1 in \ref{fig:8_config3_DimerFcs}(a) are in opposite phase each other. As a consequence of that, the orbital torques in \ref{fig:9_config3_DimerTqs}(a) show a similar property. Moreover, the presence of many dipole moments in the interaction between the spheres reduces the absolute values of both the tangential forces and the orbital torques. 

Finally, we shall discuss unexpected torques that were found for this illumination configuration. New spin components can be seen from the panels (b)-(d) of the Fig.~\ref{fig:9_config3_DimerTqs}. Panels (b), (c) and (d) correspond to spin torques around the $x$, $y$, and $z$ axes respectively. Again, there are no spin components for the case \#1 corresponding to two interacting dipole moments. 

Notice that the spin torques found also appear in coordinated form like we observed in the first configuration of illumination, Fig.~\ref{fig:4_config1_DimerTqs}; the curves in black line and in red line are equal in value but sign-opposed, see \ref{fig:9_config3_DimerTqs}(b-c). Moreover, the curves in black line and in red line are totally equal in \ref{fig:9_config3_DimerTqs}(d). The coordination of the spins preserves the symmetry of the system. The system ``feels'' an asymmetry with respect to the $x$ and $y$ axes due to this illumination variation. However, the components of the total spin torque are identically zero but their individual components, i.e. for each particle, are not. In addition, we can see from \ref{fig:9_config3_DimerTqs}(b) that the $x$-component never vanishes for this particular wavelength, in a way consistent with the results of the Fig.~\ref{fig:4_config1_DimerTqs} when $\zeta=0$ deg. Remarkably, this component of the spin torque reaches the largest values. The $x$- and $y$- components reach larger values than the $z$-component. It could be supposed that the $z$-component should ``perceive'' the variation of the polarization more directly than the rest of the components and then it would be expected to be the largest in variation but this is not the case. 

Now the ``gear'' mechanism of the light scattering already appears in 2D results, see Refs.~\cite{AbrahamE2016,AbrahamE2018_Si,AbrahamE2018_Ag}. In the present three-dimensional case, this mechanism appears for the torques along the coordinates $x$ and $y$. In the Refs.~\cite{AbrahamE2016,AbrahamE2018_Si,AbrahamE2018_Ag} this kind of induction appears in only one dimension, here analogous to what we found along the $x$ direction, i.e. spin rotation with respect to the $x$-axis. Now the new component, namely the $y$-spin, makes the particles to rotate with respect to the $y$-axis in ``counter rotation'' each other.

It is worth to mention that the phases of the curves of the orbital torque and $z$-spin torques are opposite for \#1189. This has sense since some spin-orbit coupling is expected. The total torque exerted on the whole system has to be zero on average because the illumination has no net angular momentum. Consequently, the electromagnetic field around the system must have a net contribution to the rate of angular momentum that is exactly compensated with the rate of angular momentum that is taken for the movement of the system. An indication of such effect was shown in Ref.~\cite{AbrahamE2016} for 2D systems of plasmonic dimers. A similar same phenomenon is expected in 3D systems (not shown here).

Note that the scaling factors of the orbital torque and the spin torques differ in the constant value $k_0d$. Although it appears that the module of the $z$-spin torque is larger than the module of the orbital torque for the scaled values, it is not. To compare all the values of the Fig.~\ref{fig:9_config3_DimerTqs} each other one must multiply the orbital torque by the adimensional factor $k_0d$ in a way that it takes the same units than the spin torques, i.e. $3V_nu_E$. In this particular case, the factor is $k_0d\approx5$. Then, the maxima of the orbital torque are $1.4$ times larger than the maxima of the $z$-component of the spin torque. However, the maxima of the $x$-component of the spin torque result larger than the maxima of the orbital torque in $\simeq2.9$ times for this case.

Note that the curve for the $x$-component of the spin torque differs in phase in $90$ deg. with respect to the curve for the $y$-component. This is a phase difference of geometrical origin because these components are orthogonal to each other.  However, the whole movement of the system results complex in general and, of course, the relative phases and modules of the torques differ when other wavelengths are observed. This analysis was made in order to illustrate the new effects regarding the dynamics of optically coupled nanoparticles. In particular, as said, the dynamics for metal or plasmonic nanoparticles will be different as the scattering properties of the system are ruled by surface resonances and not by volume resonances as in this case.

\section{Conclusions}

This paper outlines some unexpected consequences of the multiple scattering between two nanospheres when illuminated with time-harmonic fields that do not carry angular momentum. In particular, the study is focused on the optical forces and torques exerted on silica dimers and it is supported by DDA calculations. Three illumination configurations were explored, including a spectral study of the mechanical inductions as it was presented in previous works for 2D systems. Unusual spin torques were found for illumination with plane waves having a linear polarization. The results were explained in terms of asymmetries in the inner fields of the spheres. 

The variation of the induced forces with the angles of illumination were also studied. Remarkably, new degrees of freedom were predicted for the movement of the dimer. It was found that the whole dimer can deviate from the direction expected by the radiation pressure. When the polarization of the illumination was varied, new spin torques were found that would respond to a spin-orbit coupling. Orbital torques are coordinated with the spins of the particles in this case. If the illumination is symmetric with respect to the dimer, the spins appear in a coordinated form in such a way that the net torque for the dimer results in zero. 

In the present case of dielectric nanosphere dimers, the fields are a consequence of the coupling of morphology-dependent resonances on the spheres. However, the results of this work are supported by previous works; the new movements predicted here are of general validity. They can be obtained with independence on the materials simulated and the geometry of the scatterers \cite{AbrahamE2016,AbrahamE2018_Ag,AbrahamE2018_Si,Abraham2015_2}. In the case of plasmonic systems, the new effects are a consequence of the coupling of surface plasmon resonances.

In addition, the problem of approaching the system by single dipole-dipole interactions was compared against the complete problem that involves many dipole moments for each particle. The results show clearly how the multipolar interactions between the particles are essential in considering the realistic dynamics of the system. The dimer can be represented by two dipoles as many works do for a first approximation but this simple model cannot predict many important features of the mechanics of the optically-coupled scatterers.  

\begin{acknowledgments} 
The author would like to thank Dr. Antonio Garc\'ia-Mart\'in from IMN-CSIC for sharing interesting discussions on the topic.
\end{acknowledgments}

\end{document}